\documentstyle[aps,multicol,prl,epsfig]{revtex}
\draft

\newcommand {\e}   {{\rm e}}
\newcommand {\eps} {\varepsilon}

\begin{document}

\draft
\title{Elastically-Driven Linker Aggregation between
Two Semi-Flexible Polyelectrolytes}
\author{Itamar Borukhov$^{1,2}$, Robijn F. Bruinsma$^{2,3}$,
William M. Gelbart$^{1}$ and Andrea J. Liu$^{1}$}
\address{$^1$ Department of Chemistry and Biochemistry,
University of California at Los Angeles, Los Angeles, CA 90095, USA}
\address{$^2$ Department of Physics,
University of California at Los Angeles, Los Angeles, CA 90095, USA}
\address{$^3$ Instituut-Lorentz for Theoretical Physics,
Universiteit Leiden Postbus 9506, 2300 RA Leiden, The Netherlands}
\date{\today}

\maketitle

\begin{abstract}
The behavior of mobile linkers connecting two semi-flexible charged
polymers, such as  polyvalent counterions connecting DNA or F-actin 
chains, is studied theoretically.
The chain bending rigidity induces an effective
repulsion between linkers at large distances
while the inter-chain electrostatic repulsion leads to 
an effective short range inter-linker attraction.
We find a rounded phase transition from a dilute linker gas 
where the chains form large loops between linkers 
to a dense disordered linker fluid connecting parallel chains.
The onset of chain pairing occurs within the rounded transition.
\end{abstract}


\begin{multicols}{2}
\narrowtext


Highly charged chains of the same sign can be linked together by 
mobile polyvalent ions of the opposite sign
(polyvalent counterions). 
This phenomenon has been observed for a 
wide variety of systems, including solutions of DNA 
\cite{Bloomfield,Sikorav},
F-actin \cite{Kawamura,Pollard,Janmey,Safinya} and 
polystyrene sulfonate \cite{DeLaCruz:PSS}. 
For flexible chains, polyvalent counterions
can induce collapse of the chains into a
globular compact structure \cite{DeLaCruz:PSS}; 
alternatively, one can argue that the flexible
chains mediate attractions between polyvalent counterions that
cause them to aggregate.  For rigid rods, on the other hand, the
polyvalent counterions always repel each other along the axis of the
rods, and the attraction between rods
is attributed to ion-ion correlations among rods
\cite{Oosawa,RodAttraction}.


\begin{figure}
\centerline{\epsfxsize=3.0in \epsfbox{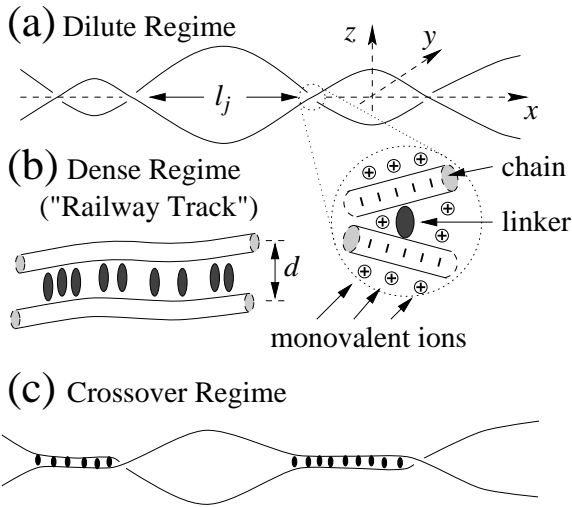}}
\vspace{\baselineskip}
\caption{\protect\footnotesize 
Schematic view of two chains connected by linkers at
(a) low linker densities (the dilute regime); 
(b) high  densities (the disordered ``railway track'' structure) 
and (c) intermediate densities (the crossover regime).
One chain is in the $y=0$ plane while the other is in the $y=d$ plane.}
\label{fig:short_scheme}
\end{figure}

In this paper, we will study the intermediate case of semi-flexible
chains, appropriate to biologically important polymers such as DNA.
It has been shown theoretically that counterion correlations can modify 
the bending rigidity of a single semi-flexible chain 
\cite{Rouzina,Golestanian,Diamant} and can even render the chain
unstable to collapse \cite{Golestanian}. 
Here, we consider two semiflexible chains from a different
point of view: instead of studying how counterions modify the
effective interactions between monomers on chains,
we examine how chain flexibility modifies the
effective interaction between generalized {\it linkers}, which could be 
simple polyvalent counterions \cite{Safinya} or weakly-binding
(crosslinking or bundling) proteins \cite{Pollard}.
Alternatively, the linkers could represent hydrogen bonds 
connecting the two strands of a DNA double helix undergoing denaturation
\cite{DNAmelting,Kafri}.
We use this effective interaction to study the 
many-body statistical mechanics of linkers. A similar approach has
been fruitful for understanding behavior of proteins that link
together elastic membranes \cite{Goulian,Becky}.

Our calculations yield three main results.  First, we find that the
chain-mediated interactions between linkers are non-monotonic.  At
large linker separations the chain bending elasticity leads to a
long-ranged repulsion, while at short distances the electrostatic
repulsion between chains leads to a short-ranged attraction between
linkers.  Consequently, there is a repulsive barrier in the interaction
between two linkers at intermediate separations.  Second,
the unusual shape of the effective potential leads to interesting
phase behavior in the many-linker system.  Since the two-chain system
is one-dimensional, there is no true phase transition 
\cite{PhaseTransition}; instead, we find a
rounded transition from a dilute phase of
linkers where the chains form large loops to a dense one where the
chains are parallel
and close together. The rounded transition is accompanied by large
fluctuations in the spacing of linkers.  
This is reminiscent of large fluctuations in the
separation of membranes near their binding/unbinding transition
\cite{Leibler}.
Third, we find that the onset of aggregation of single chains 
into pairs occurs within this rounded transition.
Our results  suggest that in many-chain systems, 
the rounded transition might be found near the onset of bundling.  


Our model consists of two charged semi-flexible chains held together
by a series of linkers (Fig.~\ref{fig:short_scheme}).
Each chain carries a negative linear charge density $\lambda=-e\xi/l_B$
where $e$ is the electronic charge, $l_B=e^2/\eps k_BT$ is the Bjerrum length,
$\eps$ is the solution dielectric constant and $k_BT$ is the thermal
energy.  
Here, $\lambda$ and $\xi$ are the effective linear charge density and 
the corresponding, dimensionless, Manning-Oosawa parameter 
\cite{Oosawa,Brenner}, respectively.
In the presence of salt at concentration $c_{b}$, the Debye-H\"uckel
screening length $\kappa^{-1}=1/\sqrt{8 \pi l_{B} c_{b}}$
characterizes the exponential decay of electrostatic interactions.
We assume that a linker of size $b$ constrains the two chains 
of radius $r_s$ at a fixed separation $d=2r_s+b$.

The first step is to calculate the effective interaction between two
linkers separated by a distance $l$.  This energy can be estimated by
considering the total energy of a periodic array of linkers with
separation distance $l$ along the $x$-axis.
For simplicity, we assume that each chain is restricted to a plane
parallel to the $xz$-plane.
One chain is located in the $y=0$ plane and follows the curve
    $\{x,y,z\}=\{x,0,z(x)\}$ where $z(\pm l/2)=0$ and $z(0)=h/2$.
The second chain is located in the $y=d$ plane and follows the curve
    $\{x,y,z\}=\{x,d,-z(x)\}$.
A periodic trial function is assumed for the two chains
\begin{eqnarray}
      z(x) =  {h\over 2} \cos\left( {\pi x/l} \right)
\label{cosine}
\end{eqnarray}
where $h$ is related to the crosslink angle $\theta$ through 
$\tan(\theta/2) = \pi h/2 l$ and will be determined variationally.
It should be emphasized that although the exact shape of the trial function
affects the numerical values it should not have a qualitative effect on the
results.

The electrostatic energy per linker is given by
\begin{equation}\label{Eel}
	E_{\rm el} = {\xi^2 \over l_B}
	\int_{-l/2}^{\l/2} {\rm d}x  \left(1+z_x^2    \right)^{1/2}
	\int_\infty^\infty {\rm d}x' \left(1+z_{x'}^2 \right)^{1/2}
	{\e^{-\kappa r_{12}}\over r_{12}}
\end{equation}
where $ r_{12}^2 = (x-x')^2 + d^2 + (z(x)+z(x'))^2 $ and 
$z_x \equiv {\rm d}z/{\rm d}x$.
In Eq.~\ref{Eel} and in the following discussion all energies
are expressed in terms of the thermal energy $k_BT$.
Note that if two straight chains cross at an
angle $\theta$, the electrostatic energy is \cite{Brenner}
\begin{equation}
E_{\rm el} = {2\pi \xi^2 \e^{-\kappa d} \over
		\kappa l_B\sin\theta} \equiv
	    {\Gamma \over \sin\theta}
\end{equation}
favoring a crossing angle of $\theta=90^{\circ}$ 
and providing a basic energy scale in the problem.  

This interaction competes with the bending energy
\cite{PerHansen}
\begin{equation}
	E_{\rm bend} = l_p
	\int_{-l/2}^{l/2} {\rm d}x  \left(1+z_x^2 \right)^{-5/2}
	z_{xx}^2
\end{equation}
which favors small crossing angles.
The chain persistence length $l_p$ characterizes the decay
of correlations in the chain orientation.
Any dependence of bending rigidity on the composition of the surrounding 
solution \cite{Rouzina,Golestanian,Diamant,OSF} is implicit 
in $l_p$.

\begin{figure}[tbh]
\centerline{\epsfxsize=3.0in \epsfbox{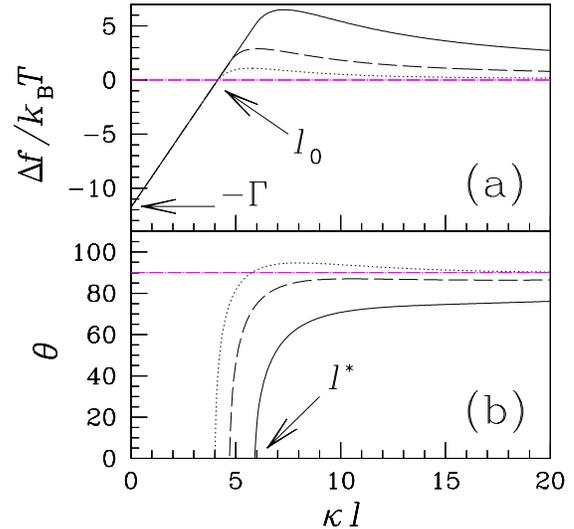} }
\caption{\protect\footnotesize 
(a) Interaction free energy per crosslink 
 $\Delta f(l)=$ $ f(l)-\Gamma$ and (b) crossing angle $\theta$ 
as functions of the reduced distance $\kappa l$ between linkers.  
Note that $\Delta f(l)$ changes sign near $l_{0}$ (see text), 
and that $\theta$ becomes nonzero at $l^{*}$ (denoted by an
arrow for the solid curve).
The curves correspond to different values of the chain persistence length 
 $l_p=10$\AA\ (dots), 
 $l_p=50$\AA\ (long dashes) and
 $l_p=200$\AA\ (solid curve).
We use $d=25$\AA, $\xi=4$, $\kappa^{-1}=10$\AA\ and $l_B=7$\AA.}
\label{fig:short_df} 
\end{figure}

The total free energy per linker $f(l;h)=E_{\rm el}+E_{\rm bend}$
can be calculated numerically and minimized with respect to
$h$ \cite{longpaper}.
Typical interaction energies $\Delta f(l)\equiv f(l)-f(\infty)$,
as well as the behavior of the crossing angle $\theta$,
are depicted in Fig.~\ref{fig:short_df}. 
Note that $f(l\to\infty) = \Gamma$.


The competition between inter-chain electrostatic repulsion and 
intra-chain bending rigidity
leads to an effective attraction between linkers at short separations
and an effective repulsion between linkers at large separations.
At short separations, the bending energy dominates
over the electrostatic energy and the chains remain parallel to each
other (crossing angle $\theta=0$).  Since the chains are
straight, the bending energy does not otherwise affect the linker
interaction.  Furthermore, the total electrostatic repulsion between chains is
fixed and does not depend on the position of the linkers.  As the
density of linkers increases, the energy per linker
decreases. The effective interaction 
can be estimated from the electrostatic energy of two 
parallel chains \cite{Brenner}. It is attractive and depends
linearly on the linker separation:
$ \Delta f_{\rm short} \simeq -\Gamma (l_0-l)/l_0 $
where $\kappa l_0=\pi/\e^{\kappa d}K_0(\kappa d)$ and
$K_0$ is the zero-order modified Bessel function.
The origin of the
attraction can also be understood by considering only two
linkers: by joining to form a single junction instead of two
separate junctions, the electrostatic repulsion between the chains is
reduced by roughly the junction energy $\Gamma$.
Thus, the
inter-chain {\it repulsion} leads to an effective inter-linker
{\it attraction}.

At very large inter-linker separations, the electrostatic repulsion
dominates the bending energy and the crossing angles saturate to
$\theta=90^{\circ}$. Since the chains are perpendicular at the
junctions, the electrostatic repulsion does not otherwise affect the
interaction between linkers.
Compressing the linkers together costs bending energy, leading to 
a long-range repulsion of the form
$ \Delta f_{\rm long} \simeq {\alpha l_p/l} $, where $\alpha$ is a
constant of order unity.  We estimate $\alpha\simeq\pi/\sqrt{2}$ by
assuming constant curvature
of the two chains in between junctions 
and minimizing with respect to $\theta$.
The crossover between the two regimes occurs at a separation $l^*$
where $| \Delta f_{\rm long}|\simeq|\Delta f_{\rm short}|$.  For
$l_{p} \alt \Gamma l_{0}$ (as is the case 
for the parameters of Fig.~\ref{fig:short_df}) we obtain
$ l^* \simeq l_0 + {\alpha l_p/\Gamma} $.  For stiffer chains, we
find $\l^{*} \approx \sqrt{\alpha l_{p} l_{0}/\Gamma}$.

The final inter-linker potential we use is
\begin{eqnarray}
	v(l)=\Delta f(l) + c\ln(l/l_p)
\end{eqnarray}
where the second term
estimates the entropy loss of two random walks of step size $l_p$ 
constrained to cross after the same number of steps at a distance $l$ 
from each other \cite{DNAmelting}.
We use $c=3/2$ which corresponds to ideal chains in three dimensions. 
A recent estimate which includes excluded volume interactions between 
different parts of the chains yields $c\simeq 2.1-2.2$ 
\cite{Kafri}.
This term is relevant only at large distances $l\gg l_p$.  
In addition, the linker
size $b$ provides a lower cut-off on the inter-linker separation
$l$.  This corresponds to a hard-core repulsion 
or, in the case of DNA denaturation, to the size of one base pair.
Other direct
interactions between linkers such as the Coulomb repulsion between
polyvalent counterions can also be included \cite{longpaper} but do not
affect our main results.


With the effective linker interaction now in hand, we can study a
one-dimensional fluid of $N$ interacting linkers located along a
straight line of length $L$ at $0\le x_1<x_2<\ldots<x_N\le L $.
At low densities (large separations between linkers) one might expect
a crystal of linkers even in this one-dimensional system because the
effective repulsion decays as $1/l$.  However, in this regime, the crossing
angle is $\theta=90^{\circ}$ and the chain configuration on one side
of the crossing junction is decoupled from the chain configuration on
the other side.  As a result, the linker interaction is only a nearest
neighbor interaction; that is, the interaction decays as $1/l$, but
only as far as the nearest linker.

For nearest neighbor interactions, one
can calculate the partition function directly from the inter-linker potential
$v(l_j\equiv x_{j+1}-x_j)$.  Note that at higher densities, the
crossing angle becomes less than $\theta=90^{\circ}$ and further-neighbor 
interactions develop.  Nevertheless, at those shorter separations,
the interaction no longer has the form $1/l$, so the further-neighbor
interactions cannot lead to new phase transitions, although they might
affect quantitative results.  We therefore neglect further-neighbor
interactions at all densities.

It is convenient to use the Gibbs ensemble in which the
natural variables are the number of linkers $N$ and the
one-dimensional pressure $P$ \cite{1d}.
In the thermodynamic limit the Gibbs free energy is 
$G = -N\ln{Z_1}$ where
\begin{eqnarray}
      Z_1 &=& \int_b^\infty {{\rm d}l\over\lambda_T} \exp\biggl(-v(l)-Pl\biggr)
\end{eqnarray}
and $\lambda_T$ is the thermal wave length.
The chemical potential is $\mu=-\ln{Z_1}$
and the linker density $\rho=N/L$ 
can be calculated from the average linker separation
$ {1/\rho} = \bar{l} = -\partial(\ln{Z_1})/\partial{P} $.
The fluctuations in the separations
$ \Delta l^2 \equiv \overline{\left(l-\bar{l}\right)^2}$
 can be similarly extracted.


\begin{figure}[tbh]
\centerline{\epsfxsize=3.0in \epsfbox{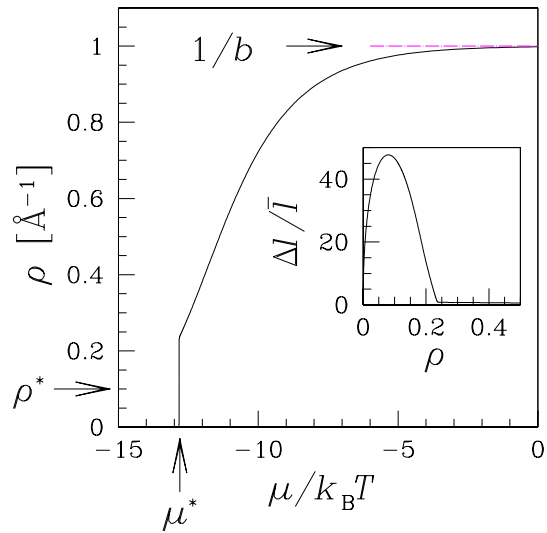}}
\caption{\protect\footnotesize 
One-dimensional linker density $\rho$ as a 
function of the chemical potential $\mu$
and (inset) the relative fluctuation in the inter-linker separation
 $\Delta l/\bar{l}$ as a function of $\rho$. 
The parameters are the same as for the solid curve 
in Fig.~\ref{fig:short_df} for which $\Gamma\simeq 11.8$; 
also, $b=1$\AA\ and $\lambda_T=1$\AA.}
\label{fig:short_gibbs}
\end{figure}

The typical behavior we obtain is shown in Fig.~\ref{fig:short_gibbs},
where we plot the linker density $\rho$ as a function of chemical
potential $\mu$ 
(shifted by $\Gamma-2\varepsilon_0$ where $\varepsilon_0$ 
is the adsorption energy of a linker onto a chain).
Note that adsorption energies and a constant $\Gamma$ term
are not included in the definition of $\mu$.
At low $\mu$ the linker density is low and the chains
form large loops between the linkers (Fig.~\ref{fig:short_scheme}a). 
At high $\rho$, on the other hand, the system resembles a
disordered railway track where the chains are rails and linkers
are ties (Fig.~\ref{fig:short_scheme}b).

Between these two limits, we find that
at a certain value of the chemical potential $\mu^*\alt-\Gamma$
the linkers aggregate together and the density jumps sharply.
The slope at the jump appears vertical because ${\rm d}\rho/{\rm d}\mu$ 
is large.
Within our model, there is no true phase transition \cite{PhaseTransition}.
However, this jump in
$\rho$ can be viewed as a rounded transition from a dilute to a
dense phase of linkers.  In the vicinity of the rounded transition
there are large fluctuations in the inter-linker separation, as shown
in the inset to Fig.~\ref{fig:short_gibbs}.
These fluctuations correspond to clusters of linkers that
increase in size as the system crosses through the rounded transition. The
crossover regime is limited to a narrow region
around $\mu^*$ but spreads over a considerable range of densities
($0\alt\rho\alt 0.2$ in the figure).
Finally, we find that as the density increases near the rounded transition,
the short-ranged
attraction between linkers sets in and the free energy drops rapidly.
As a result, paired chains become favorable compared to isolated chains.
Thus, the transition from single chains to pairs occurs near
the rounded transition ($\mu\simeq\mu^*$). 
The density $\rho^*$ at which chain pairing occurs can be estimated using 
analytical approximations for the different asymptotic 
regimes \cite{longpaper}. 
For the physical values used in Fig.~\ref{fig:short_gibbs} 
we find $\rho^*$ to be well within the rounded transition.


Our simple model can also be applied to DNA melting \cite{DNAmelting,Kafri}, 
if the hydrogen bonds are considered to be generalized linkers. 
Existing models also show a sharp melting transition.
The advantage of our approach is that some of the physical 
aspects of the chain interactions (namely, electrostatics and 
bending rigidity) are taken into account explicitly.
On the other hand, the specific double helical structure of 
double stranded DNA is not included in the model.
One straightforward generalization of the model would be to 
include a sequence-dependent linking energy reflecting 
the different number of hydrogen bonds connecting 
adenine with thymine and cytosine with guanine.

It is difficult to generalize our approach from two chains to 
many chains because many-body effects become important.
Nevertheless, we expect our results to be reflected 
in the behavior of many-chain solutions.  
We note that the ``railway track'' structure resembles a bundle where 
the chains are parallel and the linker density is high \cite{Stevens}.
Similarly, the dilute large-loop regime is reminiscent of networks where 
the chains cross at large angles and the linker density is low. Such 
structures have been observed experimentally \cite{Pollard,Safinya}.
Finally, in the limit of isolated chains the ``railway track''
is analogous to the compact torus formed in highly dilute DNA solutions
\cite{Bloomfield,Sikorav,DNAtorus}.

We thank
A. Ben-Shaul, K.-C. Lee, C. Marques, H. Schiessel, J.-L. Sikorav,
M. Stevens and J. Widom for valuable discussions.
We also benefited from correspondence with
B. Jancovici, J. L. Lebowitz and H. Spohn.
Support from NSF grant DMR-9708646 and
Israel-U.S. BSF grant 97-00205 is gratefully acknowledged.


\end{multicols}


\begin{references}
\bibitem{PhaseTransition}
The exact nature of the transition depends strongly on the interactions
between different parts of the two-chain complex. It has been argued that 
excluded volume interactions can, in fact, induce a true first order phase 
transition \cite{DNAmelting}. This effect will not be taken into account in
the current discussion.
\end{references}
\end{document}